\documentclass[twocolumn,showpacs,amsfonts,aps,prl,floatfix]{revtex4} 

\usepackage{amsmath}
\usepackage{bm}
\usepackage{graphicx}

\newcommand{\be}[1]{\begin{equation}\label{#1}}
\newcommand{\ee}{\end{equation}}

\begin{document}

\title{Viscosity from elliptic flow: the path to precision}
\date{\today}

\author{Ulrich Heinz}
\email[Correspond to\ ]{heinz@mps.ohio-state.edu}
\author{John Scott Moreland}
\author{Huichao Song}
\affiliation{Department of Physics, The Ohio State University, 
  Columbus, OH 43210-1117, USA}

\begin{abstract}
Using viscous relativistic hydrodynamics we show that systematic studies
of the impact parameter dependence of the eccentricity scaled elliptic
flow $v_2/\varepsilon$ can distinguish between different models for
the calculation of the initial source eccentricity $\varepsilon$. 
This removes the largest present uncertainty in the extraction of 
the specific viscosity of the matter created in relativistic heavy-ion 
collisions from precise elliptic flow measurements.
\end{abstract}


\maketitle

Heavy-ion collisions at the Relativistic Heavy Ion Collider produce
strongly interacting matter at extremely high energy densities 
(a quark-gluon plasma), exhibiting hydrodynamic behavior. In non-central 
collisions the collective flow is anisotropic. The degree of anisotropy,
measured by the Fourier coefficients $v_n$ of the emitted particle
distributions in the plane transverse to the beam, is sensitive to
the viscosity of the expanding fireball medium \cite{Heinz:2002rs,%
Teaney:2003kp,Romatschke:2007mq,Song:2007ux}; the largest anisotropies 
correspond to fluids with least viscosity. Measurements of the elliptic 
flow coefficient $v_2$ (which, at midrapidity, dominates all other $v_n$) 
in $200\,A$\,GeV Au+Au collisions have been compared with viscous 
relativistic hydrodynamic simulations of the fireball matter, yielding 
an upper limit for the specific shear viscosity ({\em i.e.}\ the dimensionless 
ratio between shear viscosity $\eta$ and entropy density $s$) of 
$\frac{\eta}{s} < 0.5$ \cite{Luzum:2008cw,Song:2008hj}. This is about a
factor 10 smaller than the minimal values (typically found near 
the liquid-gas transition \cite{Csernai:2006zz}) of the corresponding 
ratio measured in all other known (real) liquids \cite{Kovtun:2004de}, 
with the possible exception of strongly interacting systems of ultracold 
fermionic atoms near the unitarity limit \cite{Schafer:2007pr}. In this 
sense, the quark-gluon plasma (QGP) appears to be the most perfect liquid 
ever observed.

The uncertainty relation places a lower bound on the specific shear 
viscosity \cite{Danielewicz:1984ww}, and explicit computation in a 
large class of strongly coupled field theories, using the AdS/CFT 
correspondence, puts it near $\left(\frac{\eta}{s}
\right)_\mathrm{KSS}{\,=\,}\frac{1}{4\pi}{\,\approx\,}0.08$ 
\cite{Kovtun:2004de}. The above empirical bound is sufficiently 
close to this fundamental limit to have generated widespread interest in 
a precise measurement of the specific shear viscosity of the matter 
created at RHIC. Clearly, a measured value that saturates the KSS bound 
would have broad ramifications. It thus came as an unwelcome surprise when 
it was realized that our insufficiently precise knowledge of the initial 
fireball eccentricity $\varepsilon{\,=\,}\frac{\langle y^2{-}x^2\rangle}
{\langle y^2{+}x^2\rangle}$ \cite{Hirano:2005xf,Drescher:2006pi} (which 
depends on rapidity and, through anisotropic pressure gradients, drives 
the flow anisotropy) introduces a large, apparently irreducible  
uncertainty into the extraction of the specific shear viscosity that, 
for $\frac{\eta}{s} \sim {\cal O}\left(\frac{1}{4\pi}\right)$, can be 
100\% or more \cite{Luzum:2008cw}.
 
Let us explain the situation in more detail. To date elliptic flow 
appears to be the observable that shows the strongest sensitivity to
shear viscosity. Two main manifestations of shear viscous effects have 
been identified: (i) Shear viscosity reduces the amount of elliptic flow 
below the value generated in an ideal fluid \cite{Heinz:2002rs,%
Teaney:2003kp,Romatschke:2007mq,Song:2007ux}; for fixed initial conditions, 
in particular for a given initial fireball eccentricity, the ``viscous 
suppression factor'' grows monotonically with $\eta/s$ 
\cite{Romatschke:2007mq,Song:2007ux,Luzum:2008cw}. (ii) When scaled by 
the initial eccentricity, the elliptic flow for fixed $\eta/s$ 
is reduced more strongly in smaller collision systems that create 
shorter-lived fireballs than for larger, longer-lived fireballs. Viscous 
effects are stronger in peripheral than in central collisions, larger in 
Cu+Cu than in Au+Au collisions, and they become weaker at higher collision 
energies \cite{Song:2008si,Luzum:2009sb}.
   
Based on observation (i), Luzum and Romatschke \cite{Luzum:2008cw} made 
a first attempt to extract $\eta/s$ from experimental elliptic flow
measurements in minimum bias Au+Au collisions at $\sqrt{s}=200\,A$\,GeV.
A key ingredient of such an analysis is the ideal fluid dynamical baseline
with which the data are compared to establish the ``viscous suppression 
factor'' that is used to extract $\eta/s$. In ideal fluid dynamics the 
elliptic flow is directly proportional to the initial fireball 
eccentricity, but the latter cannot be experimentally accessed because
there are no known probes of the reaction zone that escape directly 
from the fireball and probe only the initial state, without any 
contributions from later stages of the expansion. It must therefore be
estimated theoretically from the overlap geometry corresponding to the 
impact parameter of the collision which can be extracted from 
measurements of the final hadron multiplicity and transverse energy. 
Unfortunately, theoretical models used to calculate the initial energy 
and entropy density distributions for given impact parameters differ
by up to 30\% in the predicted source eccentricity \cite{Hirano:2005xf,%
Drescher:2006pi}. In the study \cite{Luzum:2008cw} two models were 
studied whose eccentricities $\varepsilon$ differed by about 20\%. The 
corresponding 20\% variation in the ideal fluid baseline for 
the elliptic flow $v_2$ led to variations by more than a factor 2 in 
the extracted value of $\eta/s$. While the hydrodynamical calculations 
in \cite{Luzum:2008cw} made several other assumptions (in particular in 
the equation of state) that affect the ideal fluid baseline 
for $v_2$, their combined effects are likely smaller than that produced 
by the uncertainty in $\varepsilon$. More importantly, however, it is known 
how to systematically improve on these approximations in the future and 
thus dramatically reduce their contribution to the systematic error of 
$\eta/s$. On the other hand, it appears impossible to eliminate the 
uncertainty in the source eccentricity on purely theoretical grounds, 
and it is not known whether it can ever be measured experimentally. 
With the method used in \cite{Luzum:2008cw}, a precise meaurement of 
$\eta/s$, with errors below about a factor 2, thus appears to be impossible.  

In this Letter we show that, by exploiting the observation (ii) above,
one can make measurements that permit to clearly distinguish between the 
two models for $\varepsilon$ studied in \cite{Luzum:2008cw}. This clears 
the path to a precise extraction of $\eta/s$ (or at least of a combination 
of the specific shear and bulk viscosities \cite{Song:2008hj,Song:2009je}) 
from elliptic flow measurements.    

\begin{figure}[ht]
\includegraphics[width= 0.9\linewidth,clip=]{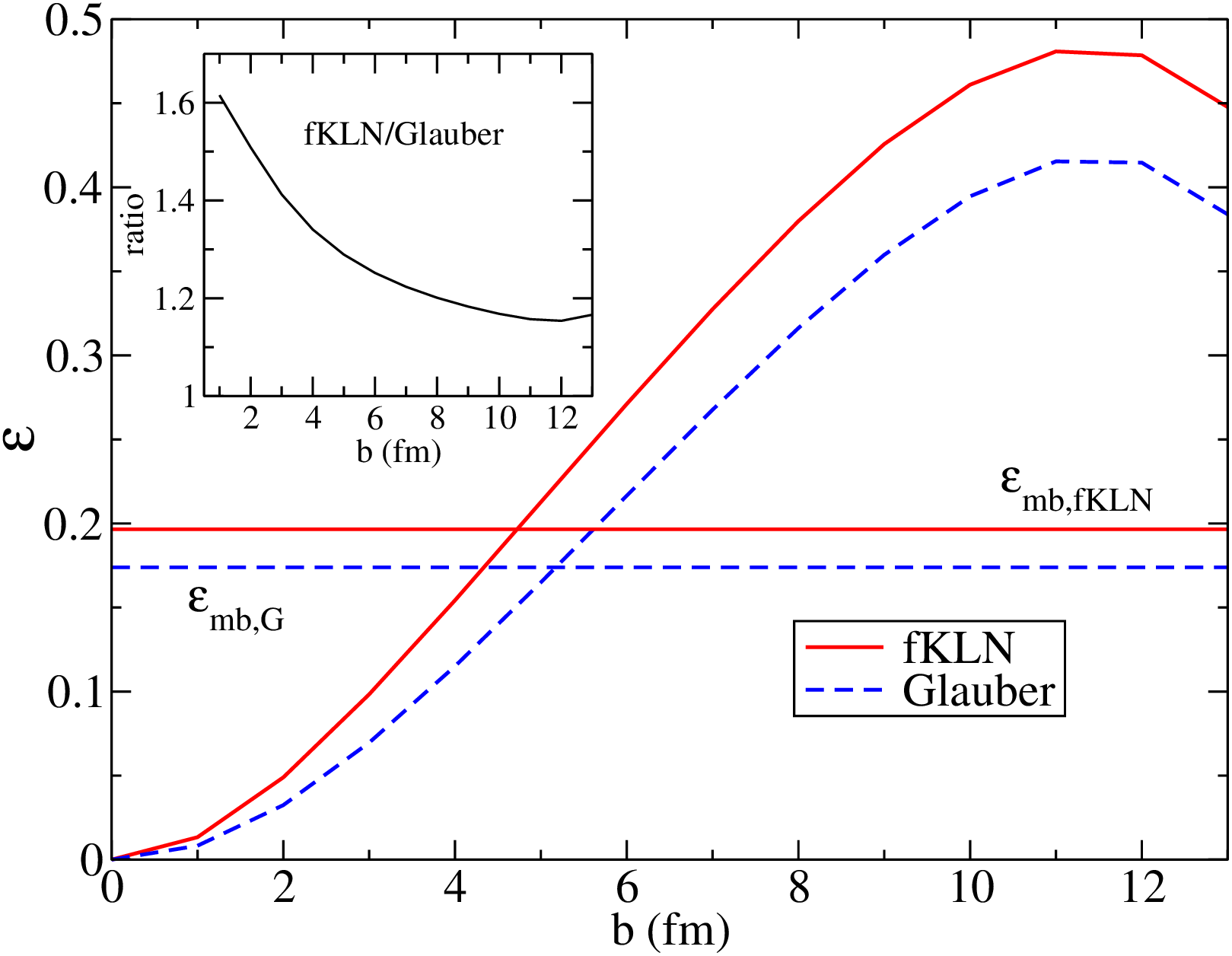}
\caption{(Color online)
Impact parameter dependence of the initial fireball eccentricity 
$\varepsilon$ (see text for details). 
}
\label{F1}
\end{figure}

Figure~\ref{F1} shows the initial spatial fireball eccentricity calculated 
from the Glauber \cite{Kolb:2001qz} and fKLN \cite{Drescher:2006pi,%
Kharzeev:2001yq} models for 200\,$A$\,GeV Au+Au collisions as a function 
of impact parameter $b$. It is defined as 
$\varepsilon(b) = \frac{\int d^2x_\perp\,(y^2{-}x^2)\,e(\bm{x}_\perp;b)}
{\int d^2x_\perp\,(y^2{+}x^2)\,e(\bm{x}_\perp;b)}$ where $e(x,y;b)$ 
is the energy density in the transverse plane at $z{\,=\,}0$; here 
$z$ denotes the beam direction, $x$ the direction of the impact parameter, 
and $y$ points orthogonal to the reaction plane. For the Glauber model 
$e(\bm{x}_\perp;b)$ is taken to be proportional to a superposition of 
wounded nucleon (85\%) and binary collision (15\%) 
densities \cite{Hirano:2005xf}. In the fKLN model the shape of 
$e(\bm{x}_\perp;b)$ is controlled by the dependence of the gluon saturation 
momentum $Q_s$ on the transverse position $\bm{x}_\perp$. We compute it 
according to Ref.~\cite{Drescher:2006pi}, using a program kindly provided 
by the authors \cite{fKLN}. 

%
%

A quick look at Fig.~\ref{F1} shows that the eccentricities from the 
fKLN model are ${\cal O}(20{-}30\%)$ larger than those from the Glauber 
model \cite{Hirano:2005xf,Drescher:2006pi}. Closer inspection reveals, 
however, that the excess depends strongly on impact parameter (see inset): 
at small $b$ the fKLN eccentricity is more than 60\% larger than the 
Glauber one whereas at large $b{\,>\,}8$\,fm the excess drops to $< 20\%$. 
Had $\varepsilon_{_\mathrm{G}}$ turned out to be simply proportional to 
$\varepsilon_{_\mathrm{fKLN}}$, the constant of proportionality (and with
it the model dependence of the initial eccentricity) could have been 
simply eliminated by forming the double ratio 
$(v_2/\varepsilon)_\mathrm{peripheral}/(v_2/\varepsilon)_\mathrm{central}$
and exploiting the system size dependence of viscous effects to extract 
$\eta/s$. The inset in Fig.~\ref{F1} shows that this will not work.

To make progress, let us next look at the sensitivity of the 
eccentricity-scaled elliptic flow $v_2/\varepsilon$ to $\eta/s$ and then 
proceed to find a way to determine which model for $\varepsilon$ should 
be used for scaling the experimentally measured elliptic flow. 
The following results are obtained from viscous hydrodynamic simulations
of $200\,A$\,GeV Au+Au collisions, with constant $\eta/s$ and standard 
initial and final conditions \cite{Song:2007ux,Song:2008si}. We comment 
on possible effects from a temperature dependence of $\eta/s$ at the end. 

\begin{figure}[ht]
\includegraphics[width = 0.9\linewidth,clip=]{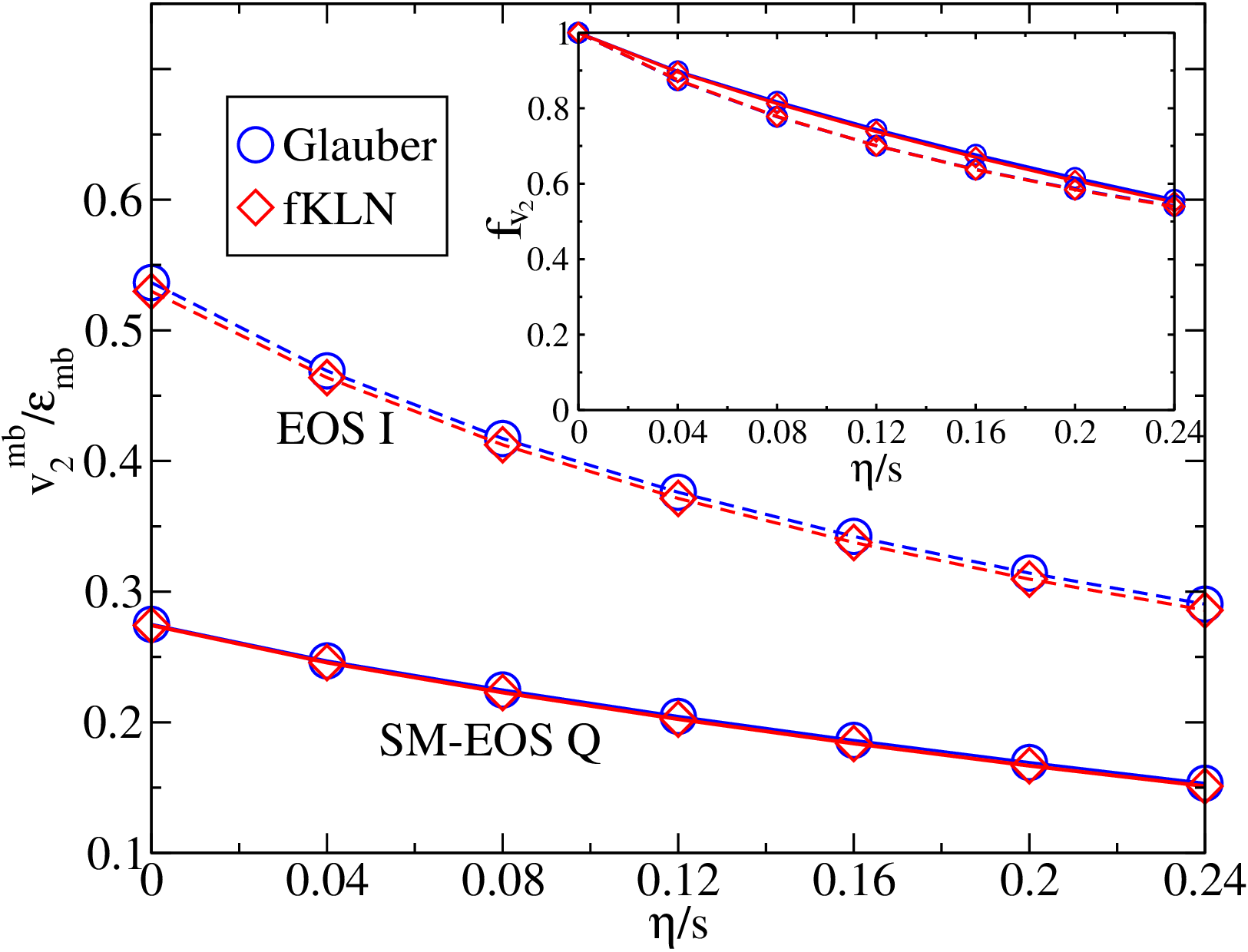}
\caption{(Color online)
Scaled elliptic flow $v_2/\varepsilon$ for minimum bias 200\,$A$\,GeV 
Au+Au collisions, as a function of specific entropy $\eta/s$, from 
viscous hydrodynamics with two different equations of state, for  
the Glauber and fKLN initial state models. 
{\sl Inset:} The fractional viscous suppression of 
$v_2^\mathrm{mb}/\varepsilon_\mathrm{mb}$ as a function of $\eta/s$.
See text for discussion.   
}
\label{F2}
\end{figure}

In Fig.~\ref{F2} we show $v_2^\mathrm{mb}/\varepsilon_\mathrm{mb}$ for 
minimum bias collisions, as a function of $\eta/s$. $v_2^\mathrm{mb}$ is 
obtained from the minimum bias pion spectrum $dN_\pi^\mathrm{mb}/(dy 
d^2p_T)$ (without resonance decay feeddown), and $\varepsilon_\mathrm{mb}$ 
is computed as above from the minimum bias energy density 
$e_\mathrm{mb}(\bm{x}_\perp)= 2\int_0^{b_\mathrm{max}}db\,b\,e
(\bm{x}_\perp;b)/b^2_\mathrm{max}$, with $b_\mathrm{max}{\,=\,}13$\,fm. 

%
%

The figure shows that {\em for minimum bias collisions} eccentricity 
scaling works in viscous hydrodynamics, {\it i.e.}\ one obtains almost 
identical curves for different initial eccentricity models (the same 
does not hold at fixed impact parameters, see below). For any given 
viscosity $\eta/s$, the scaled elliptic flow depends only on the 
stiffness on the equation of state: EOS~I, which describes a massless
parton gas with $e=3p$ and sound speed $c_s=1/\sqrt{3}$, gives more
elliptic flow per eccentricity than the softer SM-EOS~Q \cite{Kolb:1999it}
which matches a massless parton gas to a hadron resonance gas at 
$T_c=164$\,MeV through a first-order phase transition, with $c_s=0$ in 
the mixed phase and $c_s\approx1/\sqrt{6}$ in the hadronic phase. 
However, the fractional suppression of $v_2/\varepsilon$ by shear viscosity
below its ideal fluid value is almost independent of the EOS: defining
$f_{v_{2}} = \frac{\left(v_2^\mathrm{mb}/\varepsilon_\mathrm{mb}
                   \right)_\mathrm{viscous}} 
                  {\left(v_2^\mathrm{mb}/\varepsilon_\mathrm{mb}
                   \right)_\mathrm{ideal}}$ 
the fraction of scaled elliptic flow generated in viscous hydrodynamics 
relative to the ideal fluid value, the inset in Fig.~\ref{F2} shows
that this fraction is an approximately universal function of the shear 
viscosity $\eta/s$, with practically no sensitivity to the initial
eccentricity and only weak dependence on the EOS. (The somewhat lower
fractions for EOS~I are caused by earlier decoupling in this highly
explosive case \cite{Song:2007ux} which cuts off the evolution of 
$v_2$ before full saturation \cite{Kolb:1999it}.) Since the EOS can in 
principle be obtained from Lattice QCD with arbitrary precision, the 
EOS dependence is not a concern. What counts is that there is no large 
dependence on the model for the initial eccentricity $\varepsilon$ 
which can not be reliably calculated from first principles. 

Figure~\ref{F2} states that if we know $v_2^\mathrm{mb}/
\varepsilon_\mathrm{mb}$ and the EOS, we can determine $\eta/s$. But 
experimentally one can only measure $v_2^\mathrm{mb}$ while 
$\varepsilon_\mathrm{mb}$ must be calculated from a model. How can we tell 
which model for $\epsilon$ is correct? The minimum bias eccentricities for 
the fKLN and Glauber models 
($\varepsilon_\mathrm{mb}^\mathrm{fKLN}{\,=\,}0.197$ vs. 
$\varepsilon_\mathrm{mb}^\mathrm{G}{\,=\,}0.174$, see horizontal lines in 
Fig.~\ref{F1}) differ by 12\%, and Fig.~\ref{F2} shows that the 
corresponding $\sim 12\%$ uncertainty in $v_2/\varepsilon$ leads to a 
factor $\sim 2$ uncertainty in $\eta/s$ when $\eta/s={\cal O}(1/4\pi)$.  

%
\begin{figure}[ht]
\includegraphics[width = \linewidth,clip=]{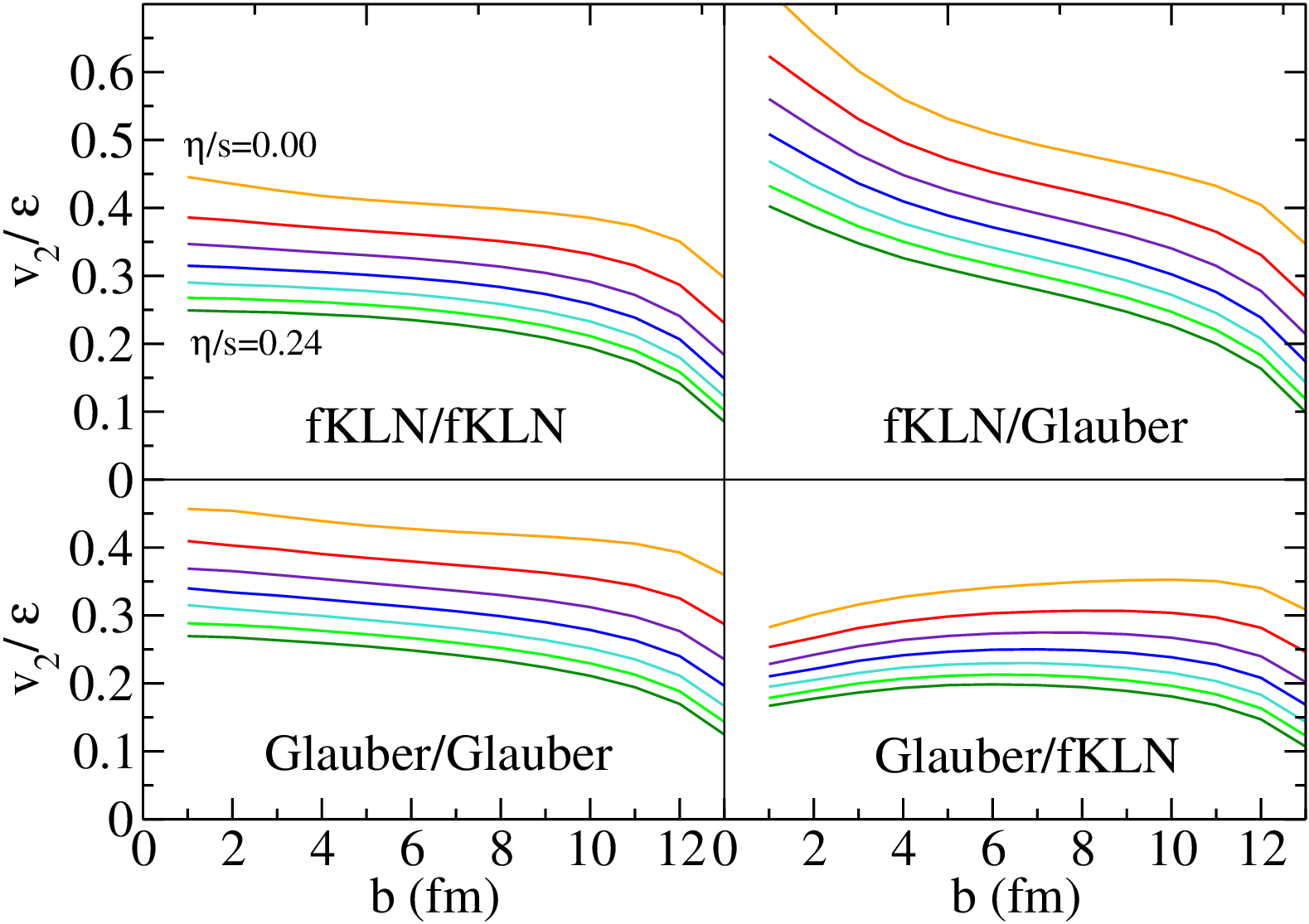}
\caption{(Color online)
Scaled elliptic flow $(v_2/\varepsilon)(b)$ from viscous hydrodynamics 
with EOS~I and either Glauber or fKLN initial conditions. The lines 
correspond to $\eta/s=0.0,\, 0.04,\, \dots,\, 0.24$ (from top to bottom in 
steps of 0.04). The left panels show 
$v_2^\mathrm{fKLN}/\varepsilon_\mathrm{fKLN}$ (top) and 
$v_2^\mathrm{G}/\varepsilon_\mathrm{G}$ (bottom),
while the right panels show the ``swapped ratios'' 
$v_2^\mathrm{fKLN}/\varepsilon_\mathrm{G}$ (top) and
$v_2^\mathrm{G}/\varepsilon_\mathrm{fKLN}$ (bottom). 
See text for discussion.
\label{F3}
}
\end{figure}
%

\begin{figure}[ht]
\includegraphics[width = \linewidth,clip=]{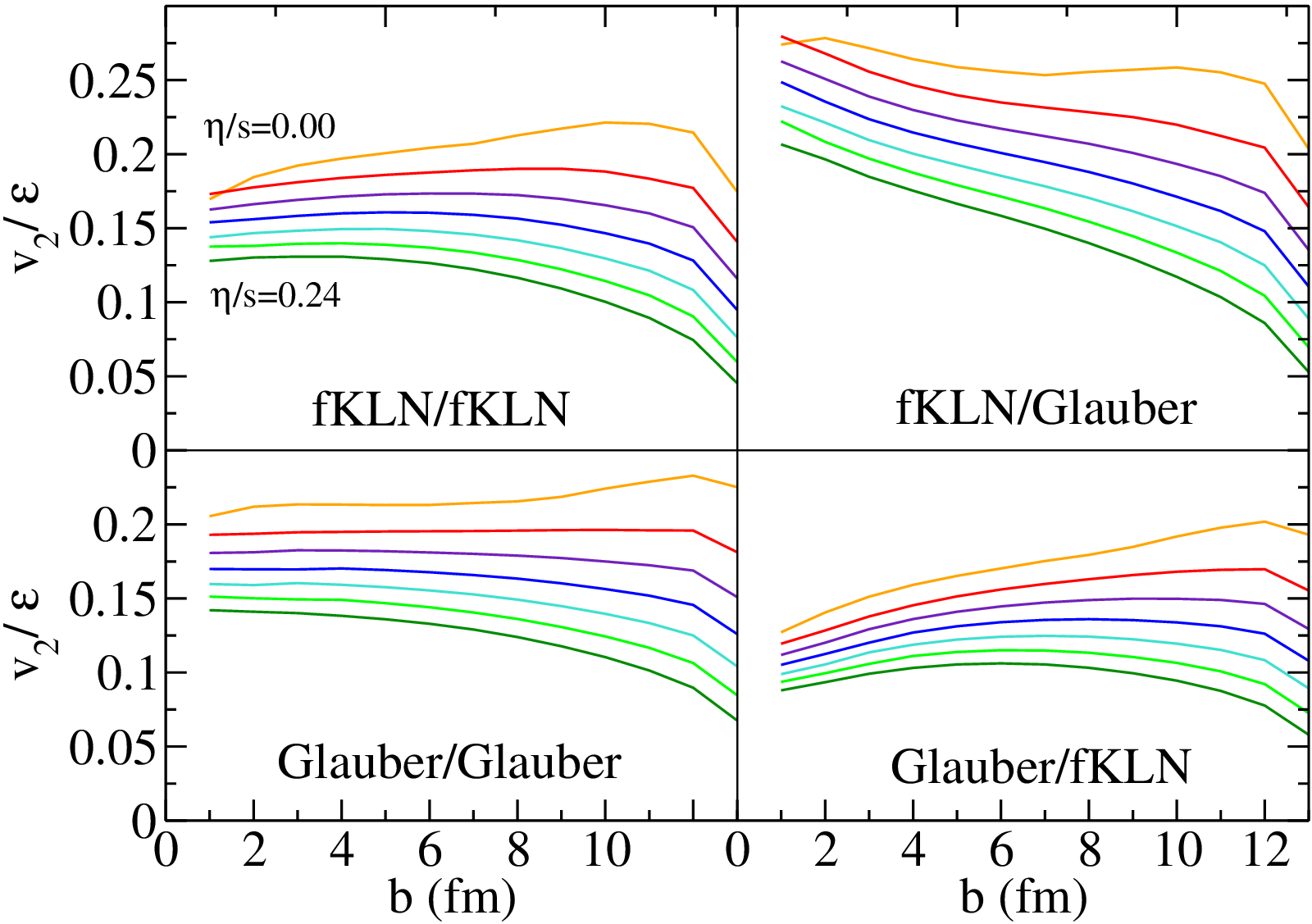}
\caption{(Color online)
Same as Fig.~\ref{F3}, but for the equation of state SM-EOS~Q with a 
quark-hadron phase transition.
}
\label{F4}
\end{figure}

The solution to this problem is given in Figs.~\ref{F3} and \ref{F4} which, 
instead of the minimum bias ratios, explore the impact parameter dependence
of $v_2/\varepsilon$ for EOS~I (Fig.~\ref{F3}) and SM-EOS~Q (Fig.~\ref{F4}). 
The left panels in these Figures show $v_2^\mathrm{fKLN}(b)/
\varepsilon_\mathrm{fKLN}(b)$ (top) and $v_2^\mathrm{G}(b)/
\varepsilon_\mathrm{G}(b)$ (bottom), respectively. Comparing the upper and 
lower left panels in Figs.~\ref{F3} or \ref{F4} one observes approximate 
eccentricity scaling at all impact parameters, although not with the same 
degree of precision as for the minimum bias average shown in Fig.~\ref{F2}. 
At each impact parameter, the viscous suppression of $v_2/\varepsilon$ is 
grows monotonically with $\eta/s$. 

For EOS~I (Fig.~\ref{F3}), $v_2/\varepsilon$ decreases monotonically with $b$, 
reflecting the earlier freeze-out and a lower degree of saturation of the 
elliptic flow in more peripheral collisions. For SM-EOS~Q we observe a more 
complex pattern: as seen in the left panels of Fig.~\ref{F4}, 
$v_2/\varepsilon$ {\em increases} with $b$ for the {\em ideal fluid} (except 
for very large $b$ where early freeze-out again takes its toll) but 
{\em decreases} with $b$ for {\em viscous fluids} once $\eta/s$ exceeds 
about once or twice the KSS bound (depending on whether we use Glauber or 
fKLN initial conditions). The increase with $b$ seen for the ideal fluid 
is well-known \cite{Kolb:1999it} and reflects the effective stiffening of 
the EOS ({\it i.e.}\ a larger effective speed of sound) as the system 
evolves out of the very soft mixed phase (which in central Au+Au 
collisions at top RHIC energies suppresses the buildup of elliptic 
flow) into the significantly harder hadronic phase (which dominates elliptic
flow buildup in the more peripheral collisions). Shear viscosity effectively
smoothes the EOS in the transition region from a first order phase 
transition into a smooth crossover \cite{Song:2007ux}, restoring the 
monotonic decrease of $v_2(b)/\varepsilon(b)$ seen also in the left panels 
of Fig.~\ref{F3}.

%
%

The key point of this Letter is, however, made by the right panels in
Figs.~\ref{F3} and \ref{F4}. In these we show the swapped ratios that
one would obtain if Nature chose fKLN initial conditions but we as 
physicists scaled the corresponding experimentally measured elliptic 
flow $v_2^\mathrm{fKLN}$ incorrectly by dividing by the initial source 
eccentricity from the Glauber model (top right panels), or {\em vice 
versa} (bottom right panels). The qualitatively different shapes of the
curves in the upper and lower right panels of Figs.~\ref{F3} and \ref{F4}
are a direct reflection of the strong centrality dependence of the 
$\varepsilon_\mathrm{fKLN}/\varepsilon_\mathrm{G}$ eccentricity ratio shown
in the inset of Fig.~\ref{F1}. It causes the swapped ratio 
$v_2^\mathrm{fKLN}/\varepsilon_\mathrm{G}$ in the top right panel to drop 
much more steeply with increasing $b$ than either of the correctly scaled 
ratios, overcoming even the stiffening effects at large $b$ from SM-EOS~Q 
in the ideal fluid case. More importantly, it causes the other swapped
ratio $v_2^\mathrm{G}/\varepsilon_\mathrm{fKLN}$ to {\em increase} with $b$
from central to mid-peripheral ($b\sim 6{-}8$\,fm) collisions. This 
increase holds even for EOS~I, over the entire range of $\eta/s$ explored
here, but it is further strengthened at RHIC energies when using
SM-EOS~Q which effectively stiffens as $b$ increases. With a more realistic
EOS that exploits the latest lattice QCD data and replaces the first order 
transition by a smooth crossover \cite{Bazavov:2009zn} we expect a 
$b$-dependence of $v_2/\varepsilon$ that interpolates between the shapes in 
Figs.~\ref{F3} and \ref{F4}.

All available experimental data from Au+Au and Cu+Cu collisions at RHIC 
indicate that $v_2/\varepsilon$ falls monotonically from central to peripheral 
collisions, irrespective of whether one uses $\varepsilon_\mathrm{fKLN}$ or 
$\varepsilon_\mathrm{G}$ to scale the measured elliptic flow $v_2$ 
\cite{Luzum:2009sb,Alver:2006wh,Drescher:2007cd,Masui:2009pw}. The lower 
right panels in Figs.~\ref{F3} and \ref{F4} then appear to exclude the 
possibility that the measured $v_2$ arises from flow driven by Glauber 
initial conditions. Furthermore, the left panels in Fig.~\ref{F4} exclude 
the possibility that the fireball medium behaves as an inviscid ideal 
fluid. We conclude that a qualitative comparison of existing data on the 
centrality dependence of the eccentricity-scaled elliptic flow suggests 
that collective flow in heavy-ion collisions at RHIC is driven by 
fKLN-like initial conditions, and that the fireball evolves as a 
viscous fluid. Extracting the precise value of its viscosity requires 
a quantitative study that goes beyond this analysis. What our work 
provides, however, is the basis for a binary decision tree that allows 
to distinguish between the Glauber and fKLN initialization models, 
thereby eliminating the largest prevailing uncertainty from such an 
extraction.

We close with a word of caution: The present analysis assumes that
the specific shear viscosity $\eta/s$ of the fireball medium is 
independent of collision centrality and thus, by implication, 
independent of temperature. Present theoretical knowledge strongly
suggests that $\eta/s$ increases during hadronization and is significantly
larger in the late hadronic than in the early QGP phase. If the effective
$\eta/s$ (averaged over the expansion history) increases dramatically
with $b$, on account of the larger role played by the hadron phase in
the evolution of $v_2$ in peripheral collisions, it may turn the
rise of $(v_2/\varepsilon)(b)$ in the lower right panel of Fig.~\ref{F4}
into a monotonic decrease, similar to the one seen in experiment. For 
this to happen the effective $\eta/s$ would have to increase from, say, 
$1/4\pi$ in central collisions to above $3/4\pi$ at $b\sim7$\,fm. While 
we believe this to be unlikely, only an explicit calculation with a 
realistic model for the temperature dependence of $\eta/s$ (and that
also includes bulk viscosity) will allow one to definitively rule out 
this possibility.  

UH gratefully acknowledges discussions with Barbara Jacak that triggered
this study. We thank A. Dumitru and Y. Nara for providing the fKLN code 
\cite{Drescher:2006pi,fKLN}, and W. Horowitz, A. Majumder and B. M\"uller 
for a careful reading of the manuscript. This work was supported by U.S.\ 
Department of Energy under contracts DE-FG02-01ER41190. JSM gratefully 
acknowledges support through a Grilly Summer Research Scholarship and 
Undergraduate Research Fellowships from The Ohio State University.



\end{document}